\documentstyle[prl,aps]{revtex} \def\narrowtext{} \tighten\twocolumn

\begin{document}
\draft
\title{
\begin{minipage}[t]{7.0in}
\scriptsize
\begin{quote}
\leftline{{\it J. Appl. Phys.}, in press (2000)}
\raggedleft {\rm NORDITA -- 1999/58 CM}\\
\raggedleft {\rm cond-mat/9909213}
\end{quote}
\end{minipage}
\medskip
Spin Wave Theory of Double Exchange Magnets}
\author{D. I. Golosov}
\address{NORDITA, Blegdamsvej 17, DK-2100 Copenhagen \O, Denmark}
\address{%
\begin{minipage}[t]{6.0in}
\begin{abstract}
An isotropic half-metallic double exchange magnet with a direct superexchange
coupling between the localized spins is studied within the
spin-wave ($1/S$) expansion. 
The momentum dependence of the spin wave spectrum (including quantum
corrections) in the ferromagnetic phase at $T=0$ is investigated. 
Based on the calculated spin wave
spectrum of the canted state, it is shown that 
as long as the external magnetic field is not too strong, the double exchange --
superexchange competition  does not result in a stabilization of a
two-sublattice canted spin ordering.
\typeout{polish abstract}
\end{abstract}
\end{minipage}}

\maketitle
\narrowtext

The phenomenon of colossal magnetoresistance (CMR) continues to
attract considerable attention to the properties of doped manganese
oxides\cite{review}.
It is commonly accepted that the ferromagnetism observed in the CMR
compounds in a wide range of doping values $x$ is due to
conduction electron-mediated double exchange interaction
\cite{DeGennes}; the latter
in turn originates from a strong ferromagnetic Hund's rule exchange $J_H$,
coupling the spins of itinerant electrons to the core spins of ${\rm
Mn}$ ions. The
present paper is devoted to an investigation of
low-temperature spin dynamics of the double
exchange magnets. Since the magnitude of localized spins is relatively
large, $S=3/2$, an appropriate theoretical framework is provided by the
spin wave theory, which allows one to study the quantum-spin
corrections within a controlled $1/S$ expansion. 
Such an expansion for the case of double exchange ferromagnets
was constructed in Ref. \cite{magnon}, to which
the reader is also referred for a review of related
theoretical and experimental work. 

In the present paper, we treat an experimentally relevant case
when an isotropic
direct nearest-neighbour antiferromagnetic superexchange coupling,
$J$, between the 
localized spins is present in addition to the double exchange
interaction (see
Ref. \cite{We}, and references therein).
We consider the ferromagnetic phase first,  analysing the momentum
dependence of the magnon spectrum at $T=0$, including quantum corrections.
We then
proceed to investigate the situation when the ferromagnetic state is
destabilized as a consequence of double exchange -- superexchange competition.
We find that due to a spin-wave instability of de Gennes'
two-sublattice canted state \cite{DeGennes} in weak fields, this
competition does not generally result in a canted 
spin ordering.

We write the Hamiltonian in the form
\begin{eqnarray}
{\cal H}=&&-\frac{t}{2} \sum_{\langle i,j \rangle,\alpha} 
( c^\dagger_{i \alpha}c_{j \alpha} +c^\dagger_{j \alpha}c_{i
\alpha} ) -
 \frac{J_H}{2S} \sum_{i, \alpha, \beta} \vec{S}_i
\vec{\sigma}^{\alpha \beta} c^\dagger_{i\alpha} c_{i\beta}+
\nonumber \\
&&+\frac{1}{S^2}J\sum_{\langle i, j \rangle} \vec{S}_i \vec{S}_j -
\frac{1}{S} H \sum_i\left\{S_i^z+ \frac{1}{2}\sigma^z_{\alpha
\beta}c^\dagger_{i\alpha} c_{i\beta}\right\} \,.
\label{eq:Ham0}
\end{eqnarray}
Here  $c_{j \alpha}$ are the electron annihilation operators,
$\vec{S}_i$ are the operators of the core (localized) spins
located at the sites of a square (or simple cubic) lattice, and
the vector $\vec{\sigma}^{\alpha \beta}$ is composed of Pauli
matrices. The effects of magnetic field $H$ on the orbital motion of
conduction electrons are omitted. 
Throughout the paper we use units in which hopping $t$, $\hbar$, $\mu_B$,
and the lattice spacing are all equal to unity, and we consider the
$T=0$ case.

We begin with the case when the superexchange interaction is not too
strong, $J<J_{c}$ (see below), so that the ground state of the system
at large $S$ is ferromagnetic \cite{assumption}.
The carrier spectrum  is then given by
$\epsilon_{\vec{k}}^{\uparrow,\downarrow}=\epsilon_{\vec{k}} \mp 
J_H/2 \mp H/(2S)$, where $ \epsilon_{\vec{k}}=-\sum_{a=1}^d \cos k_a$ \cite{tight}, and 
$d$ is the dimensionality of the lattice. We will further assume
that the value of Hund's rule coupling, $J_H$, is sufficiently large, 
resulting in a half-metallic situation with only spin-up electrons
present in the ground state.  

In the limit of $J_H \rightarrow \infty$, the leading-order term in
the magnon spectrum acquires a Heisenberg-like form
\cite{Nagaev70,Kubohata,Furukawa96},
\begin{equation}
\omega_{\vec{p}}^{(0)}=(-2J+|E|/2d)(d+\epsilon_{\vec{p}})/S+H/S.
\label{eq:spectrum0}
\end{equation}
Here and below, $E=\int_{-d}^{\epsilon_F} \epsilon \nu(\epsilon)
d\epsilon$ is the total energy of a band with
dispersion law $\epsilon_{\vec{k}}$, filled with $x$ spinless fermions
per site, whereas $\nu(\epsilon)$ and
$\epsilon_F$ are the corresponding density of states and Fermi energy.
The deviation from the cosine (Heisenberg-like) dispersion arises in the first
quantum correction to the spin wave spectrum, which can be obtained
along the lines of Ref. \cite{magnon}:
\begin{eqnarray}
&&\omega^{(1)}_{\vec{p}}=(xJ-|E|/4d)(d+\epsilon_{\vec{p}})/S^2 + 
\label{eq:real} \\
&&+\frac{1}{4S^2}\int (\epsilon_{\vec{q}}-\epsilon_{\vec{p}+\vec{q}})
n_{\vec{q}} \left\{ \int_{\epsilon_F}^d \frac{\xi
-\epsilon_{\vec{p}+\vec{q}}} {\epsilon_{\vec{q}}-\xi} \nu(\xi) d\xi
\right\} \frac{d^dq}{(2 \pi)^d}\,.
\nonumber
\end{eqnarray} 
Here $n_{\vec{q}}$ is the Fermi distribution function for spin-up
electrons.

At finite $J_H$, the spin-wave spectrum differs from that of a
Heisenberg magnet already in  leading order in $1/S$
\cite{Nagaev70,Furukawa96}: 
\begin{equation}
\omega_{\vec{p}}^{(0)}=\frac{H-2J(d+\epsilon_{\vec{p}})}{S}+\frac{J_H}{2S} 
\int \frac{\epsilon_{\vec{q}} -
\epsilon_{\vec{p}+\vec{q}}}{\epsilon^\uparrow_{\vec{q}} -
\epsilon^\downarrow_{\vec{p}+\vec{q}}} n_{\vec{q}} \frac{d^dq}{(2\pi)^d}. 
\label{eq:spectrumJH}
\end{equation}

The spin wave linewidth at $T=0$ originates from  magnon-electron
scattering processes: the oncoming magnon changes its energy and
momentum while exciting an electron from under the Fermi surface.
The leading-order term in magnon damping, $\Gamma(\vec{p})$, is  proportional to
$S^{-3}$. The expression for $\Gamma(\vec{p})$ at finite $J_H$ (and at
$J,H \geq 0$) can be
obtained by multiplying the integrand in Eq. (10) of
Ref. \cite{magnon} by
$J_H^2/(\epsilon^\uparrow_{\vec{q}}-\epsilon^\downarrow_{\vec{p}+\vec{q}})^2$,
and using our Eq. (\ref{eq:spectrumJH}) for  $\omega_{\vec{p}}^{(0)}$.
As long as the system remains half-metallic, the behaviour of
$\Gamma(\vec{p})$ is similar to that in the $J_H
\rightarrow \infty$, $J=0$ case, studied in Ref. \cite{magnon}. 

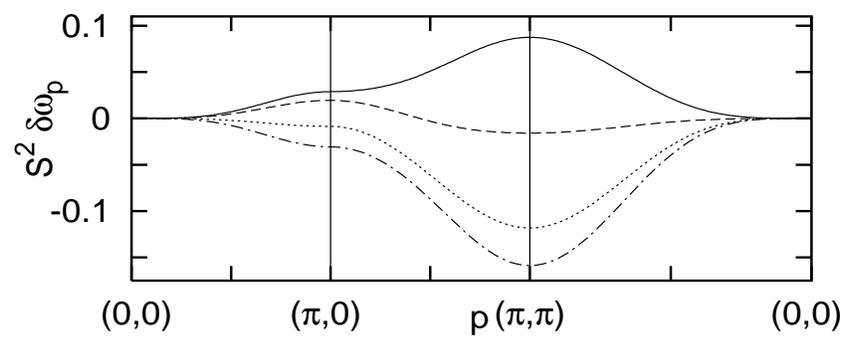
\begin{figure}
\caption{The momentum dependence of $S^2\delta \omega_{\vec{p}}$ for a 2D
system at $H=0$. The solid, dashed, dotted and dashed-dotted lines
correspond respectively to the band filling values $x=0.4$, $x=0.3$,
$x=0.2$, and $x=0.1$. }
\label{fig:delta}
\end{figure}

We will now consider the magnon spectrum in the $J_H \rightarrow
\infty$ case in more detail. At $H=0$, the leading-order term in the
deviation of the spectrum 
from a Heisenberg-like fit, $\delta \omega_{\vec{p}}=\omega_{\vec{p}}-
2D(d+\epsilon_{\vec{p}})$ (where $D$ is spin stiffness), originates
from $\omega^{(1)}_{\vec{p}}$. The momentum dependence of the 
$J$-independent quantity $\delta \omega_{\vec{p}}$ for various values
of $x$ in a 2D    
system is shown in Fig. \ref{fig:delta}. Taking also into account the
decrease of $D$ with decreasing $x$ at $J=0$ (see Ref. \cite{magnon}), we see
a strong increase of relative magnitude of $\delta \omega$ at small
$x$ \cite{Kaplan}. We also find that at small $x$, the minimum of
$\delta \omega_{\vec{p}}$ is located at $\vec{p}=\{\pi,\pi\}$, rather than at
$\vec{p}=0$ \cite{minimum}. This agrees with the results of
Refs. \cite{Nagaev72,Chubukov}, and also follows from the expression
\[
\delta \omega_{\vec{p}}=\frac{(d+\epsilon_{\vec{p}})^2}{4d^2S^2} \int 
[\epsilon_{\vec{q}}^2-(\partial \epsilon_{\vec{q}}/\partial \vec{q})^2
] \left\{\int_{\epsilon_F}^d \frac{\nu(\xi)
d\xi} {\epsilon_{\vec{q}} - \xi} \right\} \frac{n_{\vec{q}}
d^dq}{(2\pi)^d}\,,
\]
which is valid for the values of $\vec{p}$ on the Brillouin zone (BZ)
diagonal.

With increasing $J$, the ferromagnetic state becomes unstable at
$J=J_{c}$ as
indicated by a softening of the spin wave spectrum, $\omega_{\vec{p}}$, 
at a certain value of momentum,
$\vec{p}=\vec{p}_0$ (either $\omega_{\vec{p}}=0$ at
$\vec{p}=\vec{p}_0\neq 0$, or $D=0$). At $S \rightarrow 
\infty$, the entire spectrum, Eq. (\ref{eq:spectrum0}), vanishes
identically at $J=J_{c0}=|E|/4d$ \cite{metastable}. At finite $S$,
$J_c=J_{c0}+{\cal O}(S^{-1})$ and $\vec{p}_0$ may equal $\{\pi,\pi\}$
only when the latter corresponds to the
minimum of $\delta \omega_{\vec{p}}$.
In the presence of a magnetic field $H \stackrel{>}{\sim} S^{-1}$, the
softening of the spin wave spectrum at $J=J_{c0}+H/4d+{\cal O}(S^{-1})$ always
occurs at $\vec{p}_0=\{\pi,\pi\}$.

Softening of the spectrum at $\vec{p} = \{\pi,\pi\}$ is
consistent\cite{Nagaev72} with a transition into a de Gennes' canted
phase \cite{DeGennes}, in which the momenta of the two ferromagnetic
sublattices form an angle $2 \gamma$ with each other\cite{field}.
However, a definite conclusion about a transition into the canted state can not
be made until the stability of this
state is verified. As noted in Ref. \cite{We} (Eqs. (B5)
and (C4)), at $1 \gg H \geq 0$ and at sufficiently small $x$ the
canted state is 
in fact unstable with respect to single-spin fluctuations. This
statement has only  
variational significance, since it is likely that other, more
complex distortions of canted spin arrangement can lower the energy of
the system in a wider range of doping values. In order to study this
situation in a systematic way, one has to turn  to the spin wave
spectrum of the canted state, which for 
$J_H \rightarrow \infty$ and to leading order in $1/S$ is given by
a single branch, $\Omega_{\vec{p}}$, in a paramagnetic BZ:
\begin{eqnarray}
\Omega_{\vec{p}}^2&&=H(1+\epsilon_{\vec{p}}/d)(\tan^2
\gamma)X_{\vec{p}} /(2S^2)\,,\label{eq:csw1} \\
X_{\vec{p}}&&=(|E|+H)
(1-\epsilon_{\vec{p}}/d) + H(1+\epsilon_{\vec{p}}/d)/(2\sin^2 \gamma)
- \nonumber \\
&&-H+ \int {\cal P} 
\frac{\epsilon_{\vec{q}}+\epsilon_{\vec{p} + \vec{q}}}
{\epsilon_{\vec{q}}-\epsilon_{\vec{p} + \vec{q}}}
\epsilon_{\vec{q}}
n_{\vec{q}}\frac{d^dq}{(2\pi)^d}  \label{eq:csw2}
\end{eqnarray}
(see Appendix).
Here, $n_{\vec{q}}$ takes the same values as above,
and the angle $\gamma$ is given by $4dJ \cos \gamma = H +
|E|$.

The expression (\ref{eq:csw1}) vanishes at $H=0$ as a consequence of
the site-local continuous degeneracy of the canted state of a
classical ($S \rightarrow 
\infty$) double exchange -- superexchange magnet\cite{We}. This
indicates a breakdown of the $1/S$ expansion in the absence of a magnetic
field. At $H>0$, the degeneracy is lifted in favour of the canted
state, which has the highest net magnetization among the
continuously-degenerate (at $H=0$) states, and the validity of the
$1/S$ expansion is restored. 
By numerically evaluating the quantity $X_{\vec{p}}$ for $\vec{p}$
directed along the BZ diagonal and equal to the corresponding diameter
of the Fermi surface, we found the canted state to be unstable
($\Omega^2_{\vec{p}}<0$) in both 2D and 3D \cite{instab1D}, provided
that the magnetic 
field is small in comparison with the bandstructure energy scale, $H
\ll 1$. This result holds for all values of $x$, with an exception of a
narrow interval, $|x-0.5| \stackrel{<}{\sim} H^{1/2}$, around the
quarter-filling, $x=0.5$. Thus, away from $x=0.5$, the de Gennes'
canted state is not a true classical ground state at $H>0$. We believe
that this is also the case at $H=0$, especially given that
the small-$x$ instability with respect to local fluctuations occurs 
at $H=0$ as well, while at some of the larger $x$, due to
the softening of ferromagnetic spin-wave spectrum at $\vec{p}_0=0$ the
continuous ferromagnetic-to-canted transition at $J=J_c$ is impossible.

Thus, the emerging situation at $H=0$ is that the
energies of the entire manifold of classically degenerate states,
which includes the canted state, 
do not correspond to the classical energy minimum\cite{caution}. 
We note the difference from a Kagom\'{e} antiferromagnet, 
which is an example of a local continuous 
degeneracy of a {\it true} classical ground state \cite{Kagome}.

The local degeneracy of the canted state is also lifted at $H=0$ if one allows
for a finite value of $J_H \gg 1$. In  the ferromagnetic phase, the softening
of the magnon spectrum, Eq. (\ref{eq:spectrumJH}), at $J=J_{c0}+{\cal
O}(J_H^{-1})$ takes
place away from $\vec{p}=\{\pi,\pi\}$ for 
$0.5<x<1$, when $\int [\epsilon^2_{\vec{k}}- (\partial
\epsilon_{\vec{k}}/\partial \vec{k})^2] n_{\vec{k}} d^dk <0$. 
A continuous transition to the canted state is then impossible.   
At  $x<0.5$, the spin wave spectrum
in the canted state 
shows an
instability of precisely the same kind as in the $H>0$, $J_H
\rightarrow \infty$ 
case\cite{spectrum}.

The conclusion that the canted state is unstable at small $H$ is in line
with earlier theoretical results for $T=0$ \cite{canted}, and with the fact
that a checkerboard canted spin arrangement has never been seen
experimentally in the metallic phase of the CMR compounds.
Thermal fluctuations, which were taken into account within 
a single-site mean field approach in Ref. \cite{We}, were also found
to destabilize the canted spin ordering.

The author takes  pleasure in thanking A. V. Chubukov, A. Luther, and
J. Paaske
for enlightening and motivating discussions.

\appendix
\section{}

The magnon spectrum in the canted state is obtained from the
Hamiltonian (\ref{eq:Ham0}) by performing the
Holstein -- Primakoff transformation with an appropriate rotation for
each sublattice, followed by a canonical transformation
(cf. Ref. \cite{magnon}). After somewhat cumbersome calculation, we
find, to leading order in $1/S$,
$S^2\Omega^2_{\vec{p}}=A^2_{\vec{p}}-B^2_{\vec{p}}$, where 
\begin{eqnarray}
&&A_{\vec{p}}=J(1+ \cos 2\gamma) \epsilon_{\vec{p}}+H \cos \gamma - 2 d
J \cos 2 \gamma + J_H \times\label{eq:Ap} \\
&& \times \int\!\! {\cal P}\frac{J_H( \epsilon^2_{\vec{q}}\cos 2 \gamma -
\epsilon^2 _{\vec{p}+\vec{q}})-2 \epsilon_{\vec{q}}\cos \gamma \,
(\epsilon^2_{\vec{q}} -
\epsilon^2
_{\vec{p}+\vec{q}})}{4V_{\vec{q}}(V^2_{\vec{q}}-V^2_{\vec{p}+\vec{q}})} 
\frac{\tilde{n}_{\vec{q}}d^dq}{(2\pi)^d}\,, \nonumber \\
&&B_{\vec{p}}=J(1- \cos 2\gamma) \epsilon_{\vec{p}}+ \label{eq:Bp} \\
&&\,\,\,\,\,\,\,\,+ \frac{J_H^2
\sin^2 \gamma}{2} \int \!\!{\cal P} \frac{\epsilon_{\vec{p}+\vec{q}}
\epsilon_{\vec{q}} \tilde{n}_{\vec{q}}}
{V_{\vec{q}}(V^2_{\vec{q}}-V^2_{\vec{p}+\vec{q}})} \frac{d^dq}{(2
\pi)^d}\,.
\nonumber 
\end{eqnarray}
Here, the angle $\gamma$ is found from the equation
$8dJ \cos \gamma =2H - J_H \int \tilde{n}_{\vec{q}}
\epsilon_{\vec{q}}V^{-1}_{\vec{q}} d^dq/(2 \pi)^d$, and the electron
spectrum is given by $\tilde{\epsilon}^{\uparrow,\downarrow}_{\vec{q}} = \mp
V_{\vec{q}}$ with $2V_{\vec{q}}=(J_H^2-4 \epsilon_{\vec{q}} J_H \cos
\gamma + 4 \epsilon^2_{\vec{q}})^{1/2}$. In writing
Eqs. (\ref{eq:Ap}--\ref{eq:Bp}) we again assumed that the chemical
potential lies below the bottom of the upper band, whereas the Fermi
distribution function in the lower band is denoted by $\tilde{n}_{\vec{q}}$.

The $J_H \rightarrow \infty$ limit of Eqs. (\ref{eq:Ap}--\ref{eq:Bp})
yields Eqs. (\ref{eq:csw1}--\ref{eq:csw2}). It is interesting to note 
that in this limit, the value of $\Omega^2_{\vec{p}}$ at $p\ll k_F$
is proportional to $p^2H(|E| - \epsilon_F^2
\nu(\epsilon_F)+H)$. Thus, at small $H \ll
1$, the requirement $\Omega^2_{\vec{p}}>0$ for small $p$ is
identical with the 
thermodynamic stability condition for the canted phase (see
Ref. \cite{We}, Eq. (C6)).

Finally, it is instructive to point out the connection between the spin-wave
spectrum, Eqs. (\ref{eq:csw1}--\ref{eq:csw2}), and the 
small single-spin fluctuations studied in Ref. \cite{We}. It turns out
that the energies of such fluctuations \cite{misprint} are proportional
to the quantity 
\[\int \frac{X_{\vec{p}}d^dp}{2(2\pi)^d} = |E|+ \frac{H}{4 \sin^2
\gamma} + \int_{-d}^{\epsilon_F} \epsilon^2 \nu(\epsilon)
\int_{-d}^{d} {\cal P} \frac{\nu(\xi) d \xi}{\epsilon -\xi} d \epsilon.
\]
According to Eq. (\ref{eq:csw1}), at $H>0$ this quantity may be
negative only when $\Omega_{\vec{p}}$ is imaginary at
some values of $\vec{p}$. This confirms that the instability with
respect to small single spin fluctuations always leads to a spin-wave
instability, as mentioned in the text.

\begin{references}
\bibitem{assumption} In fact, this is a well-justified
assumption (cf. Ref. \cite{magnon}).
\bibitem{metastable} The ferromagnetic state may become metastable
already at $J<J_{c0}$ (cf. Ref. \cite{We}, Fig. 6).
\bibitem{instab1D} This instability also persists in 1D, since the
integral on the r.\ h.\ s. of Eq. (\ref{eq:csw2}) diverges at $p=2k_F$
and $x \neq 0.5$.
\bibitem{caution} This is expected to hold at all values of $x$, with
a possible exception of $x=0.5$.
\bibitem{spectrum}For the values of $\vec{p}$ on the BZ diagonal, we find 
$\Omega^2_{\vec{p}} \propto J_H^{-1}S^{-2}
(\partial \epsilon_{\vec{p}}/\partial \vec{p})^2
X_{\vec{p}}\int 
[\epsilon^2_{\vec{k}}-(\partial \epsilon_{\vec{k}}/\partial
\vec{k})^2] n_{\vec{k}} d^dk$  (where
$X_{\vec{p}}$ is given by Eq. (\ref{eq:csw2}) with $H=0$)
to leading order in $1/J_H$ and
$1/S$.
\bibitem{misprint} See Ref. \cite{We}, Eq. (B5). We note a misprint on
the r.\ h.\ s. of that equation, where the three integration limits should
be changed from $\pm 2$ to $\pm 2t$ ($t=\cos \gamma$). The
value of the integral is then proportional to $\cos \gamma$.
\end{references}
\end{document}